# Electrostatic Forces Mediate Fast Association of Calmodulin and the Intrinsically Disordered Regulatory Domain of Calcineurin

Erik C. Cook, Bin Sun+, Peter Kekenes-Huskey+, and Trevor P. Creamer

Department of Molecular and Cellular Biochemistry, University of Kentucky College of Medicine

+Department of Chemistry, University of Kentucky

**\*Corresponding author**

**ABSTRACT**

Intrinsically disordered proteins (IDPs) and proteins with intrinsically disordered regions (IDRs) govern a daunting number of physiological processes. For such proteins, molecular mechanisms governing their interactions with proteins involved in signal transduction pathways remain unclear. Using the folded, calcium-loaded calmodulin (CaM) interaction with the calcineurin regulatory IDP as a prototype for IDP-mediated signal transduction events, we uncover the interplay of IDP structure and electrostatic interactions in determining the kinetics of protein-protein association. Using an array of biophysical approaches including stopped-flow and computational simulation, we quantify the relative contributions of electrostatic interactions and conformational ensemble characteristics in determining association kinetics of calmodulin (CaM) and the calcineurin regulatory domain (CaN RD). Our chief findings are that CaM/CN RD association rates are strongly dependent on ionic strength and that observed rates are largely determined by the electrostatically-driven interaction strength between CaM and the narrow CaN RD calmodulin recognition domain. These studies highlight a molecular mechanism of controlling signal transduction kinetics that may be utilized in wide-ranging signaling cascades that involve IDPs.

**INTRODUCTION**

Intrinsically disordered proteins (IDPs) and proteins with intrinsically disordered regions (IDRs) are a growing class of proteins that have no well-defined three-dimensional structure.[1,2] More than 33% of eukaryotic proteins contain an intrinsically disordered region of at least 30 residues in length.[3] IDPs and IDRs do not adopt a well-defined three-dimensional structure due to a high proportion of polar (both charged and uncharged) residues and low abundance of hydrophobic and aromatic residues.[1,2] Even though IDPs and IDRs lack a specific three-dimensional structure they have a plethora of important roles which include sites of post-translational modification, sites of protein:protein interaction (through the utilization of short linear motifs or SLiMs), and disordered linkers that physically constrain domains of a protein.[4]

Previously, it has been shown by the Pappu lab that the conformational ensemble of an IDP is influenced by the linear distribution of charged residues.[5,6] The conformational ensemble of IDPs can range from collapsed globules to extended coils. Because many IDPs contain protein-binding sites, the conformational ensemble of IDPs may determine the accessibility of such protein-binding sites and, in turn, the association rate kinetics of the IDP's binding partner. Despite association rate kinetics and IDP conformational ensembles being interconnected through the influence of charged residue distribution, there are few studies characterizing this relation. We chose to study this relationship using calmodulin (CaM) and calcineurin (CaN) as our interacting pair.

CaN is a Ser/Thr phosphatase that regulates T-cell activation, long-term potentiation, and pathological cardiac hypertrophy (Fig. 2A).[7–9] CaN is heterodimeric with a catalytic A-chain and a regulatory B-chain.[10] At low intracellular calcium concentrations ([$Ca^{2+}$]), the autoinhibitory domain (AID) of the A-chain occupies the catalytic cleft of CaN and keeps CaN inactive. As intracellular [$Ca^{2+}$] rises, four $Ca^{2+}$ ions will bind to the B-chain which results in a small increase in activity.[7] $Ca^{2+}$-loaded CaM then binds to the intrinsically disordered region of CaN known as the regulatory domain or RD. The binding of CaM to CaN results in a conformational rearrangement of the RD that results in the dissociation of the AID from the catalytic site of CaN and full activation.[7,11,12]

Here we use calcineurin (CaN) constructs and intact calmodulin (CaM) to study the electrostatically driven association rate of a disordered:folded protein pair. The contribution of electrostatic forces to the association rate kinetics and binding thermodynamics of CaN and CaM were determined for CaN constructs of varying chain length and net electrostatic charge. Our chief finding was that attractive electrostatic forces between the net positively charged CaM-binding site and net negatively charged CaM increase the association rate by 100-fold compared to the absence of electrostatic forces. Moreover, these electrostatic contributions appear to arise predominantly from CaN's narrow CaM-binding motif. We speculate that similar electrostatically-driven mechanisms may play a role in CaM association with intrinsically-ordered proteins from myriad signaling pathways.

**MATERIALS AND METHODS**

Plasmids, Peptides, Protein Expression, and Purification

Three constructs were tested to determine the electrostatic and steric influences of the interaction between CaM and CaN; pCaN, the regulatory domain (RD), and full-length CaN. Fig. 4 shows the sequences of pCaN and the RD. These constructs differ in their chain length and net charge (Table 4.1). Further, two CaN mutants, CaN E376R, E381R and CaN E415K, E418K, were generated to determine the effect of modifying the conformational ensemble of the intrinsically disordered RD of CaN on the association rate kinetics with CaM.

The plasmid pETCaM1, which contains the human CALM1 gene and codes for full-length wild-type calmodulin, was transformed and expressed in BL-21 (DE3) competent *E. coli* (Agilent, Santa Clara, CA). Expression of CaM was induced by IPTG induction, and purified using a 2-trifluoromethyl-10aminopropyl phenothiazine-Sepharose (TAPP-Sepharose) column, which was synthesized at the Center for Structural Biology, Chemistry Core Facility at the University of Kentucky. For labeled CaM a cysteine was introduced in the D3C position using site-directed mutagenesis with a Q5 Site-Directed Mutagenesis Kit (New England Biolabs, Ipswich, MA). The purification procedure for CaM D3C was the same as for wild type CaM.

The plasmid pETagHisCN, which contains the gene for $\alpha$ CaNA (A chain) and B1 (B chain) that make up full-length calcineurin (CaN), was purchased from Addgene (Cambridge, MA). PETagHisCN plasmid was transformed into BL21 (DE3) CodonPlus

RIL competent *E. coli* (Agilent Technologies). Transformed *E. coli* containing pETagHisCN were grown in terrific broth until an OD$_{600}$ reach between 1.2 and 1.6, and induced with 1mM IPTG. Cell pellets were collected by centrifugation and lysis buffer added (20mM Tris (pH 7.5), 200mM NaCl, 1mM PMSF). Cells were lysed by sonication and cleared using centrifugation. After the cleared lysate passed through a 0.45um filter, it was applied to a Ni-NTA column (EMD Millipore, Billerica, MA). The Ni-NTA column was washed with 10 column volumes of 20mM Tris (pH 7.5), 200mM NaCl and eluted with 20mM Tris (pH 7.5), 200mM NaCl, and 250mM imidazole. CaN was purified further by calmodulin (CaM)-sepharose chromatography (GE Healthcare, Piscataway, NJ).

For the expression of the Regulatory Domain (RD) of calcineurin, a gene coding for residues 408-488 of Calcineurin Alpha Isoform 1 (NCBI NP_000935) was synthesized by GeneScript (Piscataway, NJ) and subcloned into the pET303 (N-terminal His6 Tag) vector (Life Technologies, Grand Island, NY). The RD plasmid was cotransformed with pETCaMI into BL-21 (DE3) (Agilent Tech., Santa Clara, CA.) *E. coli* for protein expression. Cells were grown in TB until an OD$_{600nm}$ of ~1.2-1.6 was reached and induced with 1mM IPTG. The RD was purified as stated previously.[11,12,15]

A peptide encompassing the CaM-binding domain (residues 391-414 from calcineurin A) was synthesized by Atlantic Peptides (Lewisburg, PA) and will be referred to as pCaN. The sequence of pCaN is CGARKEVIRNKIRAIGKMARVFSVLRGGW with a C-terminal tryptophan included for peptide concentration determination. An N-terminal cysteine was included for future experiments where the peptide will need to be labeled with a thiol-reactive fluorescent dye. A glycine linker was included between the cysteine and tryptophan to reduce structural effects from the added residues.

CaMKI peptide used for ITC control experiment has the sequence identity AKSKWKQAFNATAVVRHMRKLQ and was obtained from Atlantic Peptides (Lewisburg, PA).

Labeling of CaM D3C with Acrylodan

CaM D3C was labeled with acrylodan (6-Acryloyl-2Dimethylaminonaphtalene) (Life Technologies, Grand Island, NY) using thiol-modification chemistry. 200uM Acrylodan dye was combined with 100uM CaM D3C in the presence of 6M urea and 2mM EGTA, and incubated for 24 hours at 4°C. Acrylodan-labeled CaM (CaM-Acr) was exhaustively dialyzed in the presence of 6M urea to remove unconjugated dye. Finally, CaM-Acr was dialyzed into 20mM HEPES (pH 7.5). The concentration of label bound to CaM was determined using UV/Vis spectroscopy by using the extinction coefficient at 391nm which is 20,000$M^{-1}cm^{-1}$.[16] The concentration of CaM was determined using UV/Vis spectroscropy and bicinchoninic acid assay (Pierce/ThermoFisher, Grand Island, NY). Labeling efficiency CaM-Acr ranged from 70-99%.

Circular Dichroism (CD)

Circular Dichroism (CD) was performed at 37°C with a Jasco J-810 attached with a Peltier temperature controller. CD spectra were collected at a scanning speed of 20nm/min and were the average of 4 accumulations. The concentration of peptide/protein in each solution was 20uM. Solutions were allowed to temperature equilibrate for 15 minutes prior to starting a scan. All solutions contained 20mM HEPES (pH 7.5) and 10mM $CaCl_2$, and the NaCl concentration was either 0.25M, 1M, or 2M.

Isothermal Titration Calorimetry (ITC)

ITC measurements were made on a MicroCal VP-ITC (Malvern Instruments). Jacket temperature was set to 37°C (310.15K) for all ITC experiments. CaM, CaM-Acr, and pCaMKI peptide were buffer exchanged into 20mM HEPES (pH 7.5), 250mM NaCl, and 10mM $CaCl_2$ using a Bio-Rad P-2 column (Bio-Rad, Hercules, CA). Twenty-five injections of 52-120uM CaM or CaM-Acr was titrated into 4.5-10uM pCaMKI. Injections were 10uL each and took place over 10 seconds. ITC data was analyzed in Origin (OriginLab) were the heats obtained from injecting CaM ( or CaM-Acr) into buffer were substracted from injecting CaM (CaM-Acr) into pCaMKI. Thermodynamic parameters were determined from fitting the isotherms to a one-state binding model.

Fluorescence Spectroscopy

The $Ca^{2+}$-binding constant of each CaM lobe as a function of [NaCl] was determined using fluorescence spectroscopy described by VanScyoc, *et. al.*[17] The macromolecular $Ca^{2+}$-binding constant for the N-terminal lobe of CaM was determined from the phenylalanine fluorescence intensity (excitation: 250nm/emission: 280nm) which shown quenching as $[Ca^{2+}]$ increased. The macromolecular $Ca^{2+}$-binding constant for the C-terminal lobe of CaM was determined from the tyrosine fluorescence intensity (excitation: 277nm/emission: 320nm) which showed an increase in fluorescence intensity as $[Ca^{2+}]$ rose. Samples were equilibrated at 37°C prior to measurement. The $Ca^{2+}$-binding constant for each lobe of CaM was determined in two different buffer conditions:

1. Low salt buffer: 20mM HEPES (pH 7.5), 0.25M NaCl

2. High salt buffer: 20mM HEPES (pH 7.5), 2M NaCl

The concentration of CaM was 6µM in all samples. The fluorescence intensity of phenylalanine and tyrosine were plotted against $[Ca^{2+}]$. The fluorescence intensities plotted represent the average of 2 independent experiments.

Association Rate Kinetics of CaM-Acr and CaM-ligands

Association rates of CaM with CaM substrates (pCaN, RD, and CaN) were collected on an ISS K2 fluorometer (Champaign, IL) using the RX2000 stopped-flow apparatus obtained from Applied Photophysics (Leatherhead, United Kingdom) which has a deadtime of 8ms. The ISS K2 fluorometer and RX2000 were temperature controlled using a circulating water bath. All measurements were performed at 37°C. Excitation source was a Hg/Xe lamp with the wavelength set to 365nm. Sample emissions were filtered through a 455nm long-pass filter (Edmund Optics, Barrington, NJ). The on-rate of CaM-Acr with CaM ligands was determined by measuring the fluorescence intensity change of acrylodan as 50nM CaM-Acr was rapidly mixed with various concentrations of ligand.

Kinetic traces were analyzed using SigmaPlot and data was fit to either single or double exponential decay to determine the observed rate using:

Single exponential decay;     $y = a(1 - e^{-k_{on}x})$     (1)

Double exponential decay;   $y = a(1 - e^{-k_{on,1}x}) + b(1 - e^{-k_{on,2}x})$     (2)

where $a$ and $b$ are pre-exponential factors and $k_{on}$, $k_{on,1}$, and $k_{on,2}$ are the calculated on-rate constants.

On-rate constants were determined from the slope of the observed rate vs. the final concentration of CaM-ligand utilized. All buffers contained 20mM HEPES (pH 7.5) and 10mM $CaCl_2$. The NaCl concentration was modulated to alter the ionic strength of the solutions. The effects of ionic strength on the association rate constants can be described by the Debye-Huckel-like approximation:

$$\ln k_{on} = \ln k_{on,basal} = \left(\frac{U}{RT}\right)\frac{1}{1+\kappa a} \qquad (3)$$

where $k_{on}$ is the on-rate constant, $k_{on,basal}$ is the on-rate constant at infinite ionic strength, U is the energy of electrostatic interaction, R is the gas constant, T is temperature, $\kappa$ is the inverse of the Debye length, and $a$ is the minimal distance of approach which is set to 6Å.[18]

Ionic strength dependence on CaM:CaM-substrate associate rate

To distinguish steric and electrostatic forces in the association rates between CaM and each of our constructs we utilized the Debye-Hückel approximation (Eqn. 1). The Debye-Hückel approximation gives us two important pieces of information about an interacting protein pair; the contribution of electrostatic forces to the association rate (see U in Eqn. 1), and the association rate in the absence of electrostatic forces ($k_{on,basal}$ in Eqn. 1). Non-electrostatic forces will affect the $k_{on,basal}$ and include steric forces that can decrease the accessibility of the CaM-binding site. These contributions are reflected in the Debye-Hückel approximation:

$$\ln k_{on} = \ln k_{on,basal} + \left(\frac{U}{RT}\right)\frac{1}{1+\kappa a} \qquad (1)$$

where $k_{on}$ is the observed on-rate constant between CaM and the CaM-substrate. The $k_{on,basal}$ is the on-rate constant at infinite ionic strength and is equal to the y-intercept of ln $k_{on}$ vs. $\frac{1}{1+\kappa a}$. U is the energy of electrostatic interaction and is equal to the slope of ln $k_{on}$ vs. $\frac{1}{1+\kappa a}$. R is the gas constant, T is temperature, $\kappa$ is the inverse of the Debye length, and $a$ is the minimal distance of approach which is set to 6Å, and is an experimentally determined value.[18] The Debye length is the measure of the net electrostatic effect of a charged particle and how far those effects persist in solution. That is to say that a charged particle with a longer Debye length contributes an electrostatic force that extends farther than a particle with a shorter Debye length. The inverse of the Debye length, $\kappa$, can be determined from the temperature and the ionic strength:

$$\kappa = \sqrt{\frac{\varepsilon_r \varepsilon_0 RT}{2F^2 C_0}} \quad (2)$$

where $\varepsilon_0$ is the permittivity of free space, $\varepsilon_r$ is the dielectric constant, F is the Faraday constant, and $C_0$ is the molar concentration of electrolytes.

Simulation of diffusion-limited CaN/CaM association rates.

The N-domain (residue ID: 1-75 ) and C-Domain (residue ID: 76-147) structures of CaM were extracted from the crystal structure (PDB ID: 4q5u[15]). For CaN peptides, three peptides with varying lengths and charge distributions were considered: 1) pCaN: native binding region for CaM, 2) lpCaN: an elongated version of pCaN, and 3) lpcCaN: a site-directed mutant of lpCaN.

pCaN was directly extracted from the 4q5u PDB file, while for lpCaN, 5 residues were affixed to either end of pCaN (GATTA to the C-terminal and EESES to the N-terminal), based on the regulatory peptide's primary sequence. Lastly, lpcCaN was created based on mutating the three C-terminal glutamic acids to lysine. Amber14[46] was used for further structure preparation.

2) Brownian Dynamics Simulation (BD)

BD simulations were used to simulate the diffusional encounter of the CaM/CaN molecules, from which the association rate can be estimated. In present study, the Browndye Package[43] was used to simulate CaN peptide binding to the CaM C- and N-terminal domains. PDB2PQR[44] was first used to generate the pqr files from previously prepared structure files of CaM and CaN peptides. The generated pqr files were then passed into APBS[45] to evaluate the electrostatic potential of these structures, assuming an ionic strength from 150 mM KCl. APBS numerically solves the linearized Poisson-Boltzmann equation:

$$-\varepsilon \Delta \psi = \Sigma \rho_i q_i - \kappa^2 \psi \qquad (3)$$

where $\psi$ is the electrostatic potential, $\rho_i q_i$ is the charge distribution of fixed charge i, and $\kappa$ is the inverse Debye length. In present BD simulation, the reaction criterion was chosen to be 4 pairs of contacts with distance of contact being less than 4.5Å. 20000 single trajectory simulations for each system were conducted on 10 parallel processors using ***nam-simulation.*** The reaction rate constants were calculated using ***compute-rate-constant*** from the BrownDye package. For an upper bound on the estimate for the association rate and its sensitivity to ionic strength, we computed association rates for the terminal domains separately, assuming that both components bind independently:

$$\frac{1}{k_{on}} = \frac{1}{k_{Cterm}} + \frac{1}{k_{Nterm}} \qquad (4)$$

where the rates in the right hand side correspond to the association rates for the C and N terminal domains, respectively.

**RESULTS**

Fluorescence intensity changes can be used to track CaM:CaM-substrate interaction

Acrylodan is an environmentally sensitive dye that can be used to probe the environment of proteins and track events such as unfolding, binding, and conformational changes.[16,19–22] For this study, acrylodan was used to track the binding of CaM to CaM-substrates. This method of fluorescent tracking was adapted from Quintana et al who also used acyldoan labeled CaM to tracking binding to CaM-substrates.[20] Acylodan covalently attaches to proteins through free cysteines. Since wild-type CaM does not have any endogenous cysteines, a CaM mutant containing a cysteine had to be generated. The N-terminal region of CaM is disordered and we hypothesized that labeling this region with acrylodan would impose little to no structural effect of the rest of CaM. Therefore, CaM D3C was generated. CaM D3C was labeled with acrylodan and CaM-D3C labeled with acrylodan will be known as CaM-Acr.

Our lab has discovered a novel interaction between CaM and the distal helix of the regulatory domain (RD) of calcineurin (CaN).[11,12] The distal helix is a regulatory element that forms in the C-terminal end of the RD upon CaM binding to CaN. The distal helix has been shown previously to be critical for full CaN activation by CaM by promoting the dissociation of the AID from the active site of CaN. It has yet to be determined where on CaM the distal helix interacts, so in order to avoid potential disruption of this critical structure, the disordered N-terminal region of CaM was reasoned to be the best choice to place the acrylodan dye.

Far-UV circular dichroism (CD) was performed on CaM and the CaM:pCaN complex as well as the RD and CaN by itself to determine if any gross changes in secondary structure were occurring due to modulation in the ionic strength. $\alpha$-Helices are prominent in CaM as well as the CaM:CaM-substrate complex which is indicated by the double local minima of our CD spectra (Fig. 5A).[15,23] The ellipticity at 222nm is typically utilized to monitor changes in $\alpha$-helical content. The ellipticity at 222nm changes less than 3% for CaM and CaM-Acr as the [NaCl] changes from 0.25M to 2M (Fig. 5A, 6A). Similarly, the ellipticity at 222nm changes less than 2% for both CaM:pCaN and CaM-Acr:pCaN as the [NaCl] changes from 0.25M to 2M (Fig. 5B, 6B). The ellipticity of CaM:pCaN (both labeled and unlabeled) does not change significantly from 0.25M to 2M NaCl (Fig. 6A, B).

CD of CaN shows that the molar ellipticity at 222nm of CaN changes by 8% as the [NaCl] is adjusted from 0.25M to 2M NaCl (Fig. 5C, 7A). The ellipticity at 222nm of CaN at 0.25M and 2M NaCl were found not be significantly different from one another using two tailed t-test ($p > 0.05$). There was no detected secondary structural change in the RDs at 0 and 2M NaCl when comparing the molar ellipticity at 222nm (Fig. 7B). Collectively, our CD data indicates that no significant changes in structure were observed for any of our protein

Since labeling CaM with acrylodan introduces a large bulky hydrophobic fluorophore onto the surface, we investigated if this addition had an effect on binding to its ligands. To compare the binding of labeled and unlabeled CaM, we used isothermal titration calorimetry from which we can derive enthalpic, entropic, and affinity changes due to labeling. CaM's dissociation constant for CaN is estimated to be in the range of 0.24-28pM.[20,24] It is not possible to resolve affinities in this range using ITC because generated thermograms show a step-function around the stoichiometric ratio. Therefore, we used a peptide derived from CaMKI, a CaM ligand that interacts with CaM in a similar way to that of CaM:pCaN (Fig 8). The backbone atoms of CaM:pCaN and CaM:pCaMKI align with an RMSD of 2.6Å. Our ITC results indicate that CaM and CaM-Acr have no significant difference in the binding entropy (ΔS), binding enthalpy (ΔH), or dissociation constant ($K_D$) (Fig 9). Thus, the bulky hydrophobic fluorophore is unlikely to influence binding.

High ionic strength decreases CaM:$Ca^{2+}$ affinity

To ensure we are saturating CaM with $Ca^{2+}$ in the range of ionic strengths used in this study, we measured the effects of ionic strength on the $Ca^{2+}$-binding affinity of CaM. CaM's interactions with pCaN, RD, and CaN are dependent on CaM being in the $Ca^{2+}$-loaded state.[24] The interaction of $Ca^{2+}$ with CaM is electrostatically driven through the coordination of aspartate and glutamate residues and are necessary for $Ca^{2+}$-binding.[25] CaM has two globular lobes separated by a flexible linker, and each lobe contains a pair of EF-hand motifs. Therefore, CaM can interact with four $Ca^{2+}$. Due to the presence of the intrinsic fluorescence amino acids phenylalanine and tyrosine, the $Ca^{2+}$-binding constants of each lobe of CaM can be determined by tracking the fluorescence intensity of CaM.[17] Measurement of changes in phenylalanine fluorescence (excitation: 250nm, emission: 280nm) is indicative of the N-terminal lobe binding to $Ca^{2+}$, while changes in tyrosine fluorescence (excitation: 277nm, emission: 320nm) is indicative of the C-terminal lobe binding to $Ca^{2+}$.

Comparing our results of the $Ca^{2+}$-binding constants of CaM in the presence of either 250mM or 2M NaCl indicates that the $Ca^{2+}$-binding constants of both lobes are weakened by increasing the ionic strength (Fig. 10A,B). We see that the $Ca^{2+}$ affinity for the N-terminal lobe decreases by ~20-fold and the C-terminal lobe by ~400-fold (Table 2). This is not unexpected, given that the ionic strength plays an important role in the thermodynamics of binding $Ca^{2+}$ relative to other prominent ions. Therefore, we included 10mM $CaCl_2$ which is sufficient to saturate both lobes of CaM at the range of ionic strengths utilized when the association rates of CaM and its ligands were measured.

CaM-Acr binding to CaM-substrates results in fluorescence quenching

The fluorescence emission spectrum of CaM-Acr was measured in the presence and absence of pCaN, the RD, and CaN (Fig. 11A). The addition of either 2μM of pCaN, the RD, or CaN to 0.2μM CaM-Acr resulted in the quenching of CaM-Acr fluorescence emission. In this way, we can study the interaction between CaM and CaN from the basic units to the more complex full-length enzyme.

Tracking the fluorescence intensity of acrylodan (excitation: 365nm, emission: 490nm) and using a stopped-flow apparatus for fast mixing, observed association rates can be measured (Fig 11B). Fig 11B shows reaction traces of the mixing of CaM with CaN in the presence of either 10mM $CaCl_2$ or 10mM EGTA. EGTA chelates $Ca^{2+}$ and reduces the availability of $Ca^{2+}$. In the presence of excess $CaCl_2$, there is an observed binding reaction between CaM-Acr and CaN, while in the presence of excess EGTA this reaction is absent. This recapitulates the fact that CaM is known to bind to CaN in a $Ca^{2+}$-dependent manner.[24]

Screening of Electrostatic Interactions Decrease Association Rates

To measure the association kinetics, we utilized stopped-flow spectroscopy to mix a solution of CaM-Acr with either excess pCaN, RD, or CaN and tracked the fluorescence intensity of the acrylodan dye as a function of time. The kinetics traces of CaM-Acr binding to pCaN, RD, and CaN were found to best fit a single exponential decay model (Fig 12, Table 3).

Electrostatic forces between interacting proteins can be screened out by modulating the ionic strength of solution as shown by the Fersht group and others.[18,26–29] From the Debye-Hückel approximation, useful parameters can be derived by measuring the effects of ionic strength on the observed $k_{on}$. These include the $k_{on,basal}$, and the energy of electrostatic interaction (U) (Equation 1).[30] The $k_{on,basal}$ is the estimated $k_{on}$ at an infinite ionic strength and at which electrostatic forces between two proteins have been screened out. The energy of electrostatic interaction (U) is free energy contribution of electrostatic forces to the $k_{on}$.

We determined the association rate constant, $k_{on}$, at various ionic strengths by modulating the NaCl concentration (Fig. 13A). The $k_{on}$ of CaM-Acr and each of the three substrates decreased as the ionic strength increases. To put this data into the Debye-Hückel approximation we plotted the $\ln|k_{on}|$ vs. $1/(1+\kappa\alpha)$ (Fig 13B). The x-axis, $1/(1+\kappa\alpha)$, is determined from the inverse of the Debye length, $\kappa$, and the distance of closest approach, $\alpha$, which is set to 6Å. The $k_{on,basal}$ can be determined from the y-intercept and the energy of electrostatic interaction (U) from the slope (Table 4).

To provide structural insight into the experimentally-measured association rates, we performed BrownDye simulations of several CaN constructs with N- and C-terminal CaM fragments at appropriate ionic strengths. An upper bound on the association rate for the intact CaM protein was determined from the rates for the individual fragments via Equation. 4. Two obvious limitations of this approach is that binding events for tethered components are assumed to be independent, and we neglect the contributions of structural reorganization of the complex upon binding intact CaM. Nevertheless, our results demonstrate that N-terminal CaM/CaN association is the rate limiting step and coincides with experimentally measured trends. Namely, in Fig. 17A, we compared simulated $k_{on}$ values for the CaM/CaN peptides versus experimental measurements at varying ionic strengths. We note that our reaction criteria (see Methods) were tuned, such that our predicted $k_{on}$ at 1M was comparable to the stopped-flow value at ionic strength of 1.028M; all other simulations were performed without additional fitting. Consistent with the stopped-flow data, we note that the association rate decreases with increasing ionic strength. We find remarkable agreement between simulated and measured $k_{on}$ s at ionic strengths above 1M, whereas at around 0.5 M the $k_{on}$ s predicted from simulations is about 40% less than the experimental value. In Fig. 17A, we noticed that the computed $k_{on}$ s for lpCaN at all ionic strengths were significantly smaller (given the values as 0.89, 0.55, 0.44 and 0.41) than those of pCaN (given the values as 13.41, 6.60, 5.0 and 4.02), indicating an apparent hindering effect due to the flanking residues added to the two ends of pCaN. This trend is consistent with the experimental data, as $k_{on}$ s decreased with increasing construct lengths (pCaN > RD > CaN).

Despite the difference in the net charge of pCaN, RD, and CaN, the estimated electrostatic interaction energies (U) are similar (see Table 5). In contrast, the $k_{on,basal}$ is greatest for pCaN, with association with the RD being slightly slower possibly due to steric effects from increased chain length. The $k_{on,basal}$ for CaN:CaM-Acr is an order of magnitude slower than that for pCaN or the RD, which could be attributed to the addition of the bulky and folded portion of CaN.

The CaM-binding site is surrounded by well-conserved acidic patches (Fig. 3), which could modulate the association rate of CaM with the CaM-binding site. Given that the energy of electrostatic interaction of CaM-Acr with pCaN, the RD, and CaN was equal, it would not appear the flanking acidic patches contribute a significant electrostatic force. Would an alteration in the net charge of one of these acidic patches affect CaM binding to the CaM-binding site?

To investigate this question two mutants were generated that altered each of the acidic patches surrounding the CaM-binding site; CaN E376R, E381R and CaN E415K, E418K (Fig. 4, underlined residues). The CaN E376R, E381R mutant makes the net negatively charged patch N-terminal to the CaM-binding site less negative (Fig. 4, underlined residues in red). The CaN E415K, E418K mutants the C-terminal net negatively charged patch to a net positively charged patch (Fig. 4, underlined residues in orange). Fig. 14 shows the net charge per residues of wild-type RD, RD E376R, E381R, and RD E415K, E418K to illustrate how the mutation of positively charged residues into the RD of CaN affects its charge distribution.

The association rate of each these mutants and CaM-Acr was measured at various NaCl concentrations to derive the Debye-Hückel parameters stated above. The association of CaM-Acr with the CaN E376R, E381R yielded an energy of electrostatic interaction and a $k_{on,basal}$ that were very similar to wild-type CaN (Table 4). The energy of electrostatic interaction for the CaN E415K, E418K mutant increases 2-fold relative to the other constructs. The $k_{on,basal}$ of the CaN E415K, E418K showed a 16-fold decrease compared to CaN.

Similarly, we presented in Fig. 17B our data as $\ln(k_{on})$ vs. $1/(1 + \kappa a)$ for all experimental and computed data. For all cases, we note that $\ln(k_{on})$ is strictly increasing with respect to $1/(1 + \kappa a)$, which again indicates the increasing ionic strength (e.g. $\kappa \to \infty$) attenuates the driving force of the CaM/CaN electrostatic interaction. The calculated $k_{on,basal}$s for lpCaN and lpcCaN are comparable and are both about 10-fold smaller than the $k_{on,basal}$ of pCaN (shown in Table 5). For pCaN and lpcCaN, the electrostatic interaction energies are comparable while for lpCaN, the electrostatic interaction energy is just half of the other constructs. For the experimental data, comparable slopes were obtained irrespective of construct size. which indicates the energy of interaction ($U/kT$) is unaffected by the construct size. The simulated data, however indicates that the addition of flanking residues does in fact reduce $U/kT$ as shown in Table 5. This could be attributed, again, to the limit size of the conformational ensemble considered for BD simulations.

Nevertheless, in agreement with experiment, our data indicate that the E415K and E418K mutations mutants lead to higher slopes indicative of strongly electrostatic interactions

**DISCUSSION**

Electrostatic forces are known to help drive the association of folded protein pairs such as net positively charged barnase interacting with a net negatively charged barstar.[18] Electrostatic forces allow the association rate of barstar:barnase to reach $>5 \times 10^9 M^{-1} s^{-1}$. This rate is close to the Einstein-Smoluchowski limit which is the maximum rate of association between interacting molecules.[29] Mutation of residues important for electrostatic complementarity to nonionic or hydrophobic residues decreases the association rate between the pair. Folded proteins can have patches of negatively or positively charged residues on the surface that direct substrate binding.[31] In contrast, the class of proteins known as intrinsically disordered proteins (IDPs) have no well-defined three-dimensional structure due to the lack of hydrophobic and aromatic amino acids and a high proportion of polar and charged amino acids.[32] Thus, the influence of electrostatic forces on the association rate kinetics of IDPs is not entirely clear because the conformational space occupied by IDPs can be large.

Here we present kinetics data that show the electrostatic forces in IDPs and their contribution to association rates is complex. These studies utilize calmodulin (CaM) labeled with the environmentally sensitive dye acrylodan which is called CaM-Acr. We can measure the rate of CaM-Acr:CaM-substrate interaction by tracking the fluorescence intensity (Fig. 11A) of acrylodan, and this fluorescence intensity change is specific to Ca$^{2+}$-mediated CaM binding (Fig. 11B).

Acrylodan does not affect the structure or binding affinity of CaM to the CaM-binding site

We considered that possibility that labeling CaM with a fluorescent dye could alter its binding affinity for its substrates. The binding affinity of CaM and CaM-Acr to a CaM-substrate was determined using isothermal titration calorimetry (ITC). The binding affinity of CaM and the CaM-binding site of CaN is too high to be measured by ITC ($K_D$ = 0.24-28pM).[20,24]. We chose a CaM-substrate that interacts with CaM in a similar manner as the CaM-binding site of CaN but has an affinity in a range that can be measured by ITC. The pCaMKI peptide interacts with CaM in a similar fashion to the CaM-binding site of CaN (Fig. 8) and has an affinity in the nanomolar range.[33] We observed through ITC (Fig. 9) that neither the enthalpy nor affinity of CaM-Acr was different from wild-type CaM at binding to our test peptide pCaMKI.

Using circular dichroism (CD), we found that the secondary structure of CaM and CaM-Acr were nearly identical suggesting that the acrylodan label has little to no effect on the structure.(Fig. 5A,B, 6A,B). Also, the structure of CaM:pCaN or CaM-Acr:CaM shows no statistically significant change at different ionic strengths using a two-tailed t-test (p > 0.05). CaM is known to be a stable protein especially when bound to a substrate such as pCaN and is resistant to structural changes in response to ionic strength differences.[12,34,35] CaN showed a slight decrease in the CD signal as the ionic strength increased, but this change was not found to be significant (Fig. 7A). No significant changes in the CD spectra of the RD were observed (Fig. 7B).

High ionic strength decreases CaM's affinity for $Ca^{2+}$

The $Ca^{2+}$-binding affinity of CaM is known to be ionic strength dependent (i.e the higher the ionic strength, the lower the affinity).[36] The reason why ionic strength affects the $Ca^{2+}$-binding properties is disputed but is hypothesized to be due to either the effects of electrostatic screening from high concentrations of ions, or $Na^+$ and other ions directly competing with $Ca^{2+}$ for the $Ca^{2+}$-binding site on CaM. Because the binding of CaM to CaN is dependent on CaM being in its $Ca^{2+}$-bound state, and we wanted to study the effects of ionic strength on the association rate of CaM and CaM-substrates, we needed to control for the $Ca^{2+}$-binding affinity of CaM at different ionic strengths. To ensure we are saturating CaM with $Ca^{2+}$ we measured the $Ca^{2+}$-binding affinity of CaM at an ionic strength of 0.278M and 2.078M which are the lowest and highest ionic strengths used throughout this study.

CaM is a two-lobe protein with each lobe containing two $Ca^{2+}$-binding sites.[25] The binding of the N- and C-terminal lobes of CaM to $Ca^{2+}$ can be measured by tracking the fluorescence emission of phenylalanine and tyrosine respectively.[17] Note here that we are not measuring the affinity of $Ca^{2+}$ to each $Ca^{2+}$-binding site of CaM, but rather we are determining the macroscopic $Ca^{2+}$-binding affinity of $Ca^{2+}$ to each of lobes of CaM. The $Ca^{2+}$-binding affinity for the N- and C-terminal lobe of CaM at an ionic strength of 0.278M is similar to previous reports (Table 2, Fig. 10A).[17,37]

The C-terminal lobe of CaM was shown to bind $Ca^{2+}$ with a greater affinity than the N-terminal lobe at an ionic strength of 0.278M, but this relationship is reversed at an ionic strength of 2.03M. As mentioned above, the $Ca^{2+}$-binding lobes of CaM may be able to bind $Na^+$. Therefore, $Na^+$ may be directly competing with $Ca^{2+}$ for CaM's $Ca^{2+}$-binding site which would decrease the apparent affinity of CaM for $Ca^{2+}$. If $Na^+$ competes for the $Ca^{2+}$-binding sites on the C-terminal lobe with a greater affinity compared to the N-terminal lobe, this might explain why the C-terminal lobe of CaM has a low apparent affinity for $Ca^{2+}$ at high a [NaCl]. Another hypothesis is that the increase in the electrostatic screening from high concentrations of NaCl affects the C-terminal lobe much more so than the N-terminal lobe possibility due to different electrostatic charge contents of the two lobes of CaM (Fig. 16).

The $Ca^{2+}$ affinity of each of the lobes of CaM is lower at an ionic strength of 2.078M compared to 0.278M (Table 2, Fig. 10A,B). The $K_D$ of N- and C-terminal lobes for $Ca^{2+}$ increases by 20 and 400 fold respectively. Therefore, we included 10mM $CaCl_2$ during all of our CaM and CaM-substrate measurements to ensure CaM is fully bound to $Ca^{2+}$, and we can potentially avoid any effects due to the decrease in affinity of CaM and $Ca^{2+}$.

The CaM-binding site confers electrostatic attraction to CaM

We derived the energy of electrostatic interaction (U) and the $k_{on,basal}$ from the Debye-Hückel approximation (Equation 1). The value of U has been shown to estimate the strength of electrostatic attraction between interacting proteins.[38] The $k_{on,basal}$ refers to the basal association rate or the rate of interaction given that intermolecular electrostatic interactions are negligible. Steric forces can reduce reaction rates by increasing the non-reactive surface area of the protein in question and decreasing the accessibility of a binding site which would be apparent as a decrease in the $k_{on,basal}$. Models for protein:protein interaction predict that before two proteins form their final and relatively stable complex, they form a loosely bound "encounter complex". In this encounter complex, the protein pair is thought to bounce and slide across each other's surfaces and sample many orientations.[39] This pair might eventually find the proper orientation to form a stable and specific complex. In the encounter complex electrostatic forces assist proteins in finding the proper relative orientation in which to form a stable complex.[38] For example, a protein may have a binding site composed of a large amount of positively charged residues, and its binding partner's interface may be mostly net negatively charged. Thus, the electrostatic attraction may help these two complementary charged binding sites collide more frequently than if both sites were uncharged, thereby potentially leading to a faster association rate.

Using constructs derived from CaN (pCaN, RD, and CaN itself) we sought to determine the role of electrostatic forces in the CaM:CaM-substrate interaction. The small peptide pCaN represents our smallest unit CaM-substrate which just contains the net positively charged CaM-binding site of CaN. By including the rest of the disordered chain (in the RD) or the rest of the full-length enzyme, we can compare the effects of the added chain length, electrostatic charge, and steric hindrances to the association rates of CaM with the CaM-binding site. Previously, Quintana, et al. measured the association of a CaM-Acr construct labeled at the C75 position to full-length CaN at a single ionic strength.[20] The association traces determined from their lab fit a single exponential model that is consistent with our data. Their data showed that the association rate of CaN and CaM in 20mM MOPS (pH 7.0), 150mM KCl, and 5mM $CaCl_2$ at 25°C was $4.7 \times 10^7 M^{-1}s^{-1}$.[20] This was similar to our data collected in 20mM HEPES (pH 7.5), 250mM NaCl, and 10mM $CaCl_2$ at 37°C which was $3.7 \pm 0.3 \times 10^7 M^{-1}s^{-1}$.

Both our experimental and computational approaches indicated that the association rate of pCaN and CaM was higher across all ionic strengths tested (Fig. 13A,B and Table 4) and also had the highest $k_{on,basal}$ compared to other CaM-substrates. pCaN was our smallest CaM-substrate, so we hypothesize that this high $k_{on,basal}$ is due to the low amount of steric hindrances and a high degree of accessibility of the CaM-binding site to CaM. pCaN is a strong polyampholyte located in the dark green region of the Das-Pappu phase diagram in Fig. 15 which form hairpins, coils, or chimera. The presence of many positively charged residues that are well distributed across the sequence and the fact that the $f_+$ of pCaN almost puts it in the positively strong polyelectrolyte region suggests coil conformational ensembles. Coil conformations might be of functional significance to the CaM-binding site as they could promote exposure and fast binding to CaM.

The association rate of CaM and the RD was lower than the association of CaM and pCaN at every ionic strength. The addition of the acidic patches on either side of the CaM-binding site (RD compared to pCaN) did not appear to weaken the energy of electrostatic interaction of the substrate and CaM (Table 4, Fig. 13B). This suggests that the flanking acidic patches that might be expected to be electrostatically repelled from CaM do not appreciably influence the energetics of complex formation. The $k_{on,basal}$ of CaM and the RD was 70% that of CaM and pCaN indicating a potential steric influence. This would suggest that the added chain length of the RD decreases the accessibility of the CaM-binding site.

Further investigation of segments of RD reveals interesting predictions about the conformational ensemble of the RD. Segments of IDPs can take on conformational ensembles that are not necessarily dependent on the other segments of the IDP in question. As such, different segments of the RD may have different conformational ensembles. To get a general idea of the conformational ensemble of segments of the RD we analyzed the two halves of the RD using CIDER. Fig. 15 shows an analysis of the fraction of negatively and positively charged residues of four constructs; pCaN, RD, RDN-term, and RDC-term. RDN-term and RDC-term represent the two halves of the RD split into an equal number of residues. The N-terminal half of the RD has a higher charged residue content than the C-terminal half of the RD and also contains the complete CaM-binding site. The conformational ensemble of the RDN-term is predicted to be hairpin, coil, or chimera (which is composed of both hairpins and coils). The RDC-term is a weak polyelectrolyte which is predicted to be globular or tadpole. These predictions suggest that the N-terminal RD has a conformational ensemble that promotes exposure of the CaM-binding site, and the C-terminal end of the RD adopts a globular conformational ensemble that could lead to steric hindrance in the binding of CaM to the CaM-binding site. The C-terminal end of the RD may act similar to a bulky folded domain. Because the C-terminal region has a lower fraction of charged residues, it could decrease the $k_{on,basal}$ without significantly altering the energy of electrostatic interaction.

We have noted in previous studies that structural changes occur in the RD upon CaM binding and multiple sites of the RD interact with CaM, but the CaM-binding site accounts for most of the binding energy between CaM and the RD.[11,12] When CaM binds to the RD, two regions undergo a disorder to order transition. These are the CaM-binding site and distal helix region both of which fold into a $\alpha$-helix upon binding to CaM.[11,12] The residues of the RD that compose the distal helix are located on the C-terminal end of the RD and are predicted to be between residues 78-90 of our RD construct (Fig. 4). The CaM-binding site imparts high CaM-binding affinity in the 0.24-28pM.[20,24]. The binding affinity of distal helix to CaM is currently unknown. The distal helix has been shown to have a $T_m$ of 38°C which is far less stable than the CaM:CaM-binding site complex (Tm > 80°C).[34] These data would suggest that the most of the binding affinity of CaM to CaN is due to the CaM-binding site, not the distal helix. That is not to say that the binding of the distal helix to CaM is not important because it is known to be critical for the full activation of CaN.[12] The effects of the distal helix on the association rate of CaM and the CaM-binding site are currently unknown. Further deconvolution of the effects of the distal helix may be investigated with RD constructs that contain distal helix-disrupting mutations and our CaM-Acr tracking method.

The $k_{on,basal}$ of CaM-Acr binding to CaN is reduced (~10-fold reduction) compared to CaM-Acr binding to pCaN or the RD. CaN has a large folded part including the catalytic domain, CaNB, and the rest of the RD. If CaM interacts with a part of CaN that is outside the CaM-binding domain, then that could result in a non-productive interaction (no final CaM:CaN complex formation). Simply put, the folded parts of CaN could interfere with CaM binding to the CaM-binding site. The energy of electrostatic interaction of CaN and CaM was the same within the error of CaM binding to either pCaN or the RD. This suggests that the well-folded and net negatively charged portion of CaN does not interact with the CaM-binding site in such a way as to perturb the electrostatic attraction between CaM and the CaM-binding site. This idea is supported by previous reports that the RD does not interaction with folded parts of CaN.[11]

There is more to consider about what influences the conformational ensemble of the RD than just the electrostatic component. In full-length CaN the RD is tethered at both ends. The N-terminus of the RD is covalently attached to the B-chain binding helix. Because the autoinhibitory domain (AID) is at the C-terminal end of the RD, the C-terminus of the RD is attached to the active site of CaN. The tethering of the both ends of the RD might create limitations on its conformational ensemble. Tethering would prevent the RD from extending its end-to-end distance farther than ~52Å (the distance between the beginning and end of the RD).

The energy of electrostatic interaction (U) of CaM and CaN E415K, E418K is approximately twice that of pCaN, RD, or CaN suggesting that the added basic residues are contributing a more favorable attractive interaction between the CaM-binding site and CaM. In contrast, CaN E376R E381R did not show a significant change in the energy of electrostatic interaction compared to wild-type CaN. One explanation is that the positive charges that were introduced in the E415K, E418K mutant were much closer to the CaM-binding site (Fig. 4, underlined orange residues, Fig. 12C) than the E376R, E381R mutant. Thus, the positively charged residues K415 and K418 could be close enough to the CaM-binding site so that they are directly interacting with CaM as CaM approaches the CaM-binding site. These added positively charged residues might act to attract the incoming CaM, and once bound may form a salt bridge or hydrogen bond with the residues of CaM. Salt bridge formation could make the complex more stable and could be studied by determining the dissociation rate constant of these mutants. This might lead to a decrease in the apparent dissociation rate assuming that this additional electrostatic force results in a significant change in complex stability.

CaN E376R E381R mutations decrease but do not eliminate the net-negative charge of the acidic patch N-terminal to the CaM-binding site (Fig. 12B). But, a 2-fold reduction in the $k_{on,basal}$ of CaN E376R E381R was observed. Together with the CaN E415K, E418K, changing either of the acidic patches surrounding the CaM-binding site decreases the accessibility of the CaM-binding site for CaM. These mutants could favor a higher proportion of collapsed states in the RD's conformational ensemble.

The conformational ensemble of IDPs has been shown to be influenced by the linear distribution of the charged amino acids.[5] By reversing the charge on the acidic patch C-terminal to the CaM-binding site (in the CaN E415K, E418K mutant), we have potentially altered the conformational ensemble which could be the reason we observe a decrease in the $k_{on,basal}$. The decrease in the $k_{on,basal}$ represents decreased accessibility of the CaM-binding site by CaM. In contrast, our CaN E376R, E381R mutant altered the net charge of the RD to the same degree, but did not alter the $k_{on,basal}$ as much as E415K, E418K did. Besides the E415K, E418K mutants being closer to the CaM-binding site than the E376R, E381R mutants, the E415K E418K mutants also completely reversed the charge of the C-terminal acidic patch, whereas the E376R, E381R mutation decreased but did not reserve the N-terminal acidic patch. The E376R, E381R mutations might of perturbed the net charge of the N-terminal acidic patch enough to affect the conformational ensemble of the RD which led to a 2-fold reduction in the $k_{on,basal}$. . There is also the obvious difference that lysine residues could affect the structural ensemble of the RD differently than arginine residues. Further studies with E376K, E381K, and E415R, E418R mutants will report on the possible differences of basic residue identity.

CIDER analysis of these two mutants puts them in the same location of the Das-Pappu plot since they both have the same $f_+$ and $f_-$ as shown in Fig. 13 as "RDmut". Since these two mutants of CaN have the same net charge, and the location of these mutants are different we surmise that these two acidic patches play different roles in the conformational ensemble or accessibility of the CaM-binding site. Smaller sections of the RD were inspected to predict what sort of conformational ensembles occur. Fig 13 shows the results of sequences that represent the RD cut into 5 sections of equal amino acid length (shown as 1-5). The general trend of the fragments of the RD show that the N-terminal residues points 1 and 2, are closer to the strong polyampholyte regions show in dark green, and points 3, 4, and 5 are closer to the weak polyampholyte region in light green. Care must be taken to not over interpret these results, but this would suggest that the first 2/5 of the RD take on conformational ensembles of a strong polyampholyte (hairpin, coil, or chimera), and the last 3/5 of the RD take on conformational ensembles of a weak polyampholyte (globule or tadpole). We hypothesized that the decrease in the $k_{on,basal}$ of CaM and pCaN compared to the RD was due to a decrease in the accessibility of the CaM-binding site. The energy of electrostatic interaction of the association of CaM to pCaN or RD were the same and the kon,basal of the CaM and the RD was reduced compared to pCaN. We suspect the C-terminal end of the RD of reducing accessibility of the CaM-binding site rather than the acidic patches flanking the CaM-binding site that has a strong negative charge. The E376R, E381R and E415K, E418K mutants suggest that modifying the acidic patches flanking the CaM-binding site modifies accessibility of the CaM-binding site. If the conformational ensemble of the CaM-binding site is coil-like that confers a high degree of exposure, then modification of either the of acidic patches surrounding the CaM-binding site might lead to collapse of the CaM-binding site to more collapsed ensembles such as hairpins or globules. The E376R, E381R and E415K, E418K mutations conferred the same change in the overall net charge, but seeing as the E415, E418K mutation resulted in a greater effect on not only the electrostatic energy of interaction and the $k_{on,basal}$, we surmise that location of charge alterations is more important.

Altogether, the salt dependence of association rates yields valuable information on electrostatic and steric influences between interacting proteins. These differences in the CaM-substrates give us clues as to the accessibility of the CaM-binding site. Studying the structure of IDPs is difficult because IDPs tend to take on an ensemble of conformations ranging from extended coils to compact globules. The conformational ensemble of IDPs is at least partially determined by the $f_+$ and $f_-$.[6,14,40] In this work we sought to characterize the functional effect of perturbing the conformational ensemble of an IDP by altering the $f_+$ and $f_-$ as well as the distribution of the charged residues. The "function" in this case was the association rate constants of CaM and CaM-substrates. Our data suggests that the conformational ensemble can play a very important role in the binding of folded proteins to IDPs. Many IDPs contain a protein-binding site, and while this data should not be applied to all IDPs, it does raise an interesting observation. Despite IDPs having no persistent or stable three-dimensional structure, disturbances in their conformational ensemble can affect their function. We speculate that certain conformational ensembles might be functionally important for promoting protein:IDP interactions such as coil ensembles that expose a relatively large amount of surface area to solvent.

Accelerating protein:protein association rates through electrostatic interactions is likely to be a critical factor for intracellular processes that demand rapid and selective responses. CaN, as an example, is a $Ca^{2+}$-activated phosphatase important in long term depression and regulation of neuronal networks.[8] $Ca^{2+}$-transients in neuron signaling occur in the millisecond range, and, therefore, the electrostatic contributions to rapid association that we observe may play an important role in CaN activation and downstream signaling transduction.[41]

ACKNOWLEDGEMENTS:

PKH thanks UK for its startup support. PKH thanks XSEDE for comp support (give ref).

Figures and tables

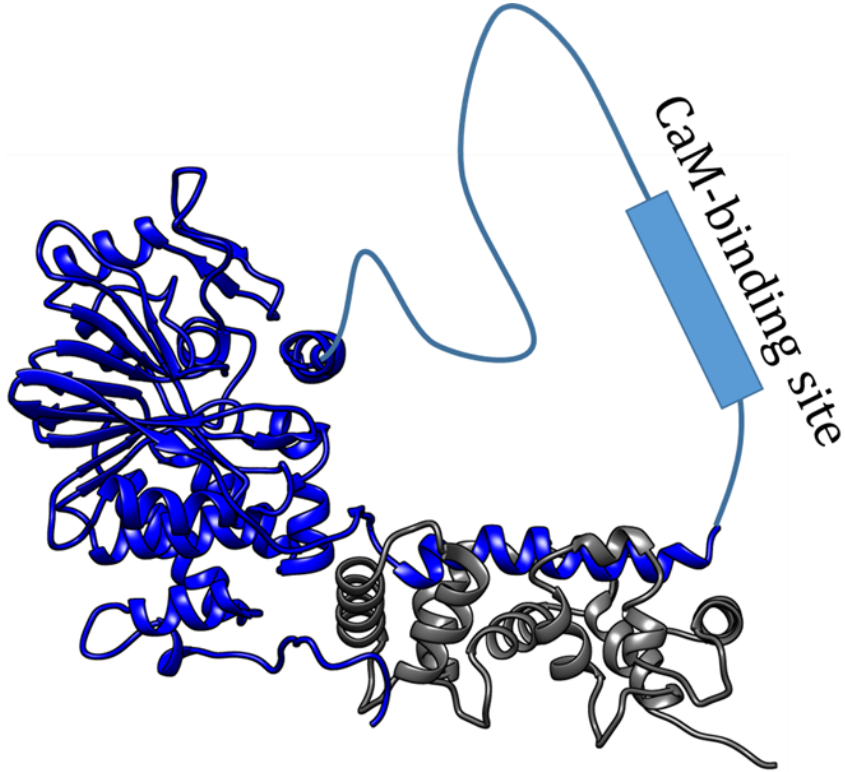
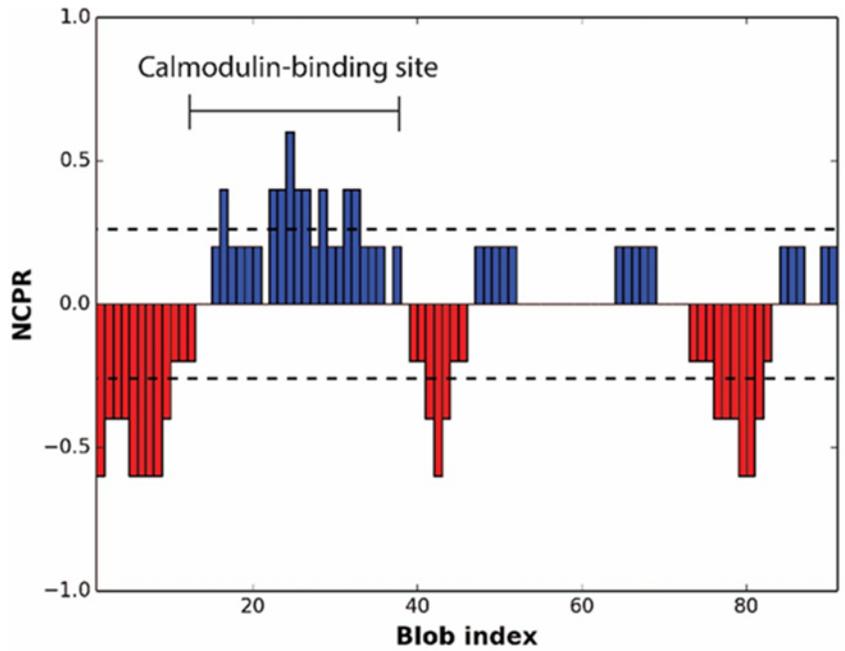

Figure 2. A. Ribbon representation of CaN B. Linear net charge per residue plot of the regulatory domain of CaN obtained using CIDER. The net charge per residue is plotted as a function of the blob index. A blob is a window of 5 residues in which the net charge per residue was determined. The net charge of this window is assigned to the center residue. CIDER is available for public use at the following URL: "http://pappulab.wustl.edu/CIDER".[5]

```
H.sapiens     YWLPNFMDVFTWSLPFVGEKVTEMLVNVLNICSDDELGSE-EDGFDG--------------
M.musculus    YWLPNFMDVFTWSLPFVGEKVTEMLVNVLNICSDDELGSE-EDGFDG--------------
G.gallus      YWLPNFMDVFTWSLPFVGEKVTEMLVNVLNICSDDELGTE-EDGFDG--------------
C.elegans     YWLPNFMDVFTWSLPFVGEKVTEMLVHILNICSDDELMAECDDTFEG--------------
S.cerevisiae  YWLPDFMDVFTWSLPFVGEKVTEMLVAILNICTEDELENDTPVIEELVGTDKKLPQAGKS
A.fumigatus   YWLPNFMDVFTWSLPFVGEKITDMLIAILNTCSKEELEDETPTSVSPS-------------

H.sapiens     -----------------ATAAARKEVIRNKIRAIGKMARVFSVLREESESVLTLKGLTPT
M.musculus    -----------------ATAAARKEVIRNKIRAIGKMARVFSVLREESESVLTLKGLTPT
G.gallus      -----------------ATAAARKEVIRNKIRAIGKMARVFSVLREESESVLTLKGLTPT
C.elegans     -----------------GVGSARKEVIRHKIRAIGKMARAFSVLREESESVLALKGLTPT
S.cerevisiae  EAAPQPATSASPKHASILDDEHRRKALRNKILAVAKVSRMYSVLREETNKVQFLKDH-NS
A.fumigatus   --AP-----SPPLPMDVESSEFKRRAIKNKILAIGRLSRVFQVLREESERVTELKTA-AG
```

Figure 3. CLUSTAL Omega alignment of the intrinsically disordered regulatory domain of orthologous calcineurins.[42] The CaM-binding site is shown in blue and has a net positive charge. On either side of the CaM-binding site are well-conserved acidic patches shown in red.

Regulatory domain (RD) construct:

```
         10         20         30         40         50
60
MAGSDDELGS EEDGFDGATA AARKEVIRNK IRAIGKMARV FSVLREESES VLTLKGLTPT
         70         80         90        100        110
GMLPSGVLSG GKQTLQSATV EAIEADEAIK GFSPQHKITG WGGGLE HHHH HH
```

pCaN peptide:

```
         10         20
CGARKEVIRN KIRAIGKMAR VFSVLRGGW
```

Figure 4. Sequence information for the RD and pCaN constructs used in this project. The RD construct is a 112-residue protein containing the entire intrinsically disordered region of CaN. The RD contains the CaM-binding site (shown in blue) surrounded by two highly conserved acidic patches (shown in red and orange). Underlined residues represent residues that were mutated to positively charged residues in full-length CaN to investigate the role of the flanking net negatively charged patches.

The RD contains a His-tag on the C-terminal end to aid in its purification. The pCaN peptide contains the net-positively charged CaM-binding site of the RD. An N-terminal cysteine and glycine linker was added to pCaN so that it can labeled with cysteine reactive fluorescence dyes for future experiments. A C-terminal tryptophan was mutated into the 101 position of the RD construct and added the C-terminus of pCaN to increase the accuracy of protein concentration determination by UV/Vis.

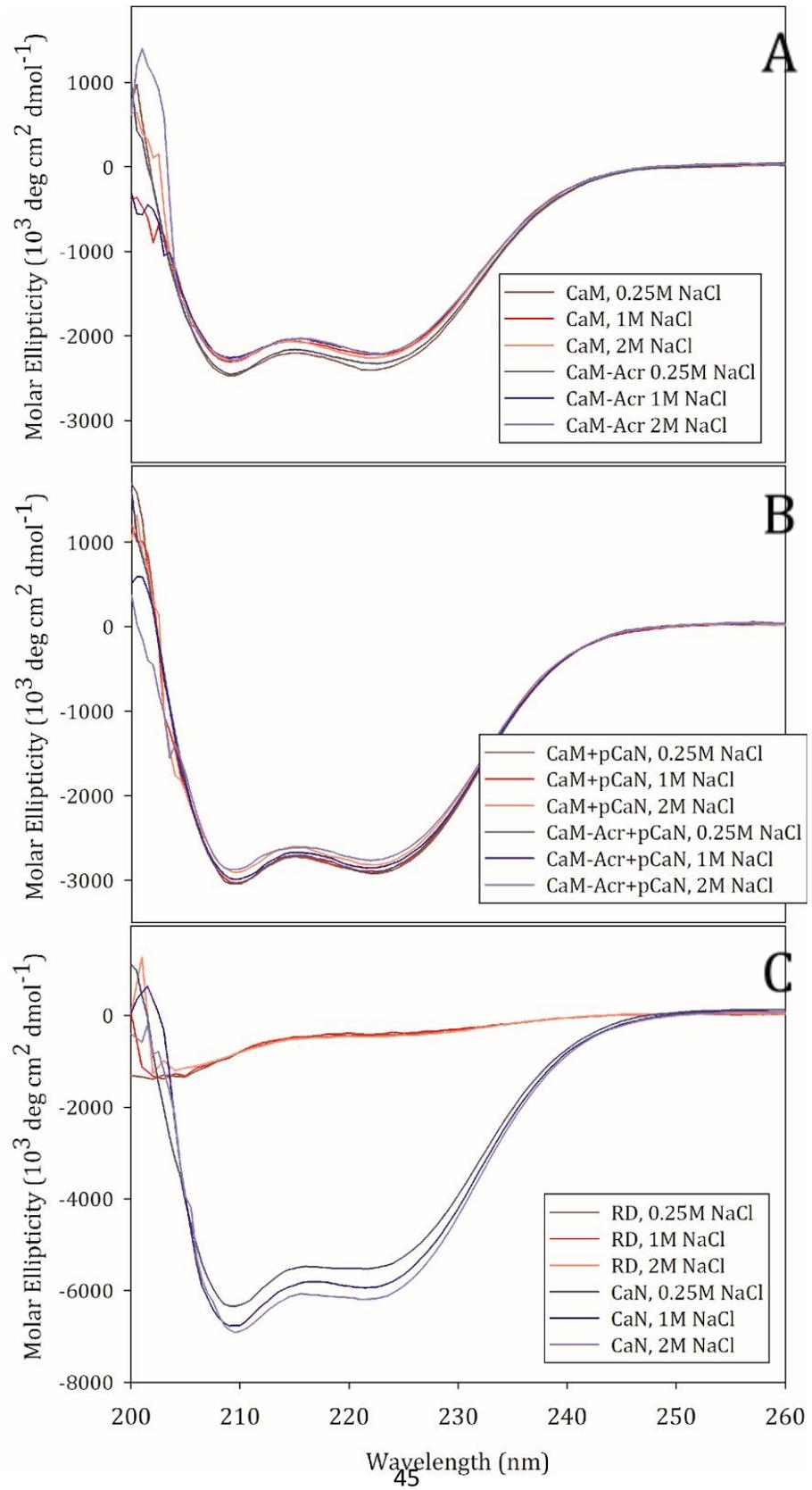

Figure 5. A and B. CD of CaM and CaM:pCaN complex in 20mM Tris (pH 7.5) and either 0.25M, 1M, or 2M NaCl at 37°C. A. Comparison of the CD of CaM and CaM-Acr in the presence of either 0.25M or 2M NaCl. B. Comparison of the CD of CaM:pCaN and CaM-Acr:pCaN in the presence of either 0.25M or 2M NaCl.

C. CD of CaN and the RD in 20mM Tris (pH 7.5) and either 0.25M, 1M, or 2M NaCl at 37°C. Signal noise was observed below 205nm for all CD spectrum with either 1M or 2M NaCl present which is due to the high concentrations of salt.

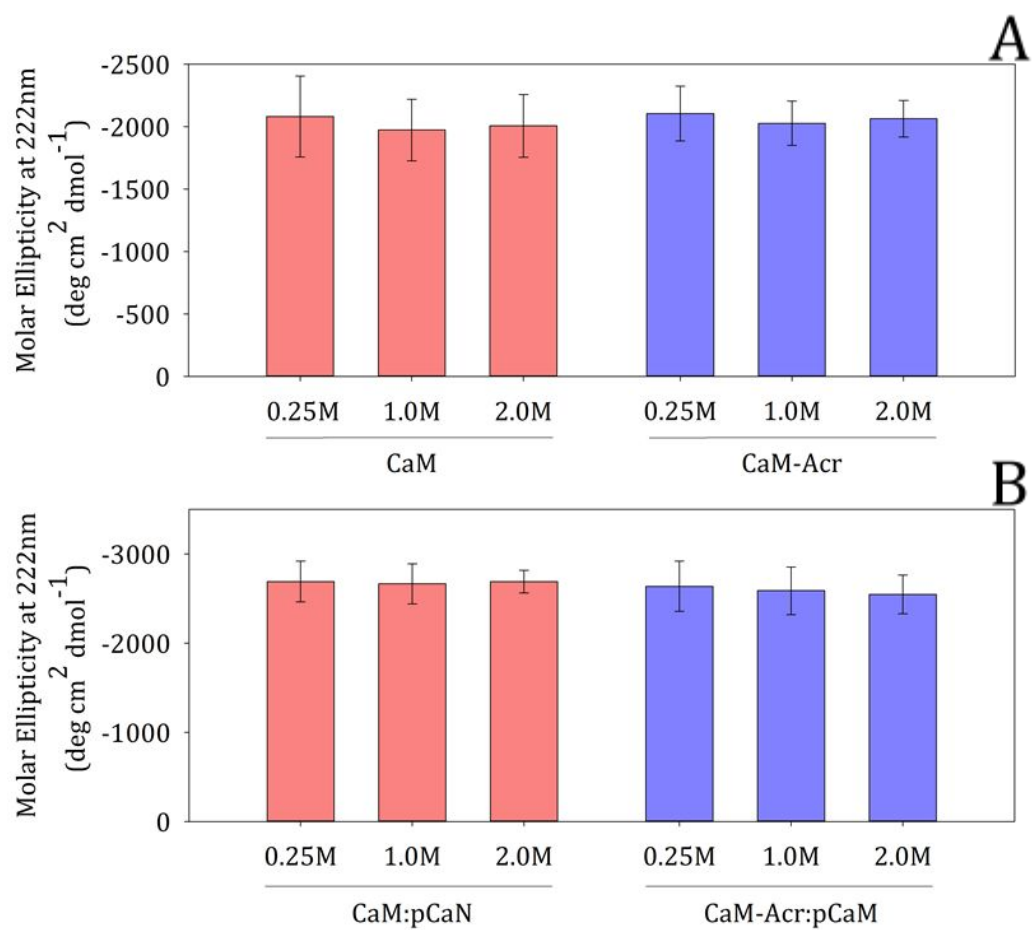

Figure 6. Molar ellipticity at 222nm of CaM and CaM-Acr in the absence (A) or presence (B) of pCaN at 0.25M, 1.0M, and 2.0M NaCl. These data include are the average of 3 independent CD experiments and error bars represent the standard deviation.

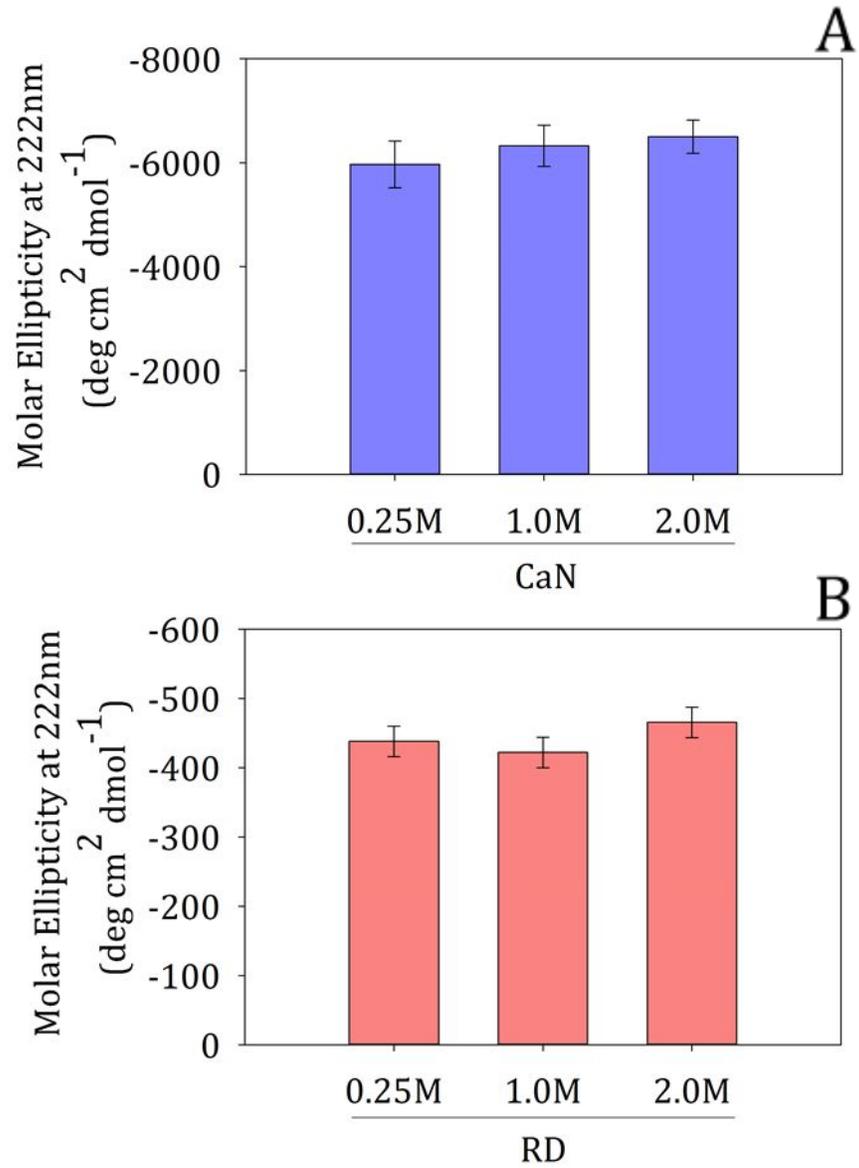

Figure 7. Molar ellipticity at 222nm of CaN (A) and the RD (B) construct at 0.25M, 1.0M, and 2.0M NaCl. These data include are the average of 3 independent CD experiments and error bars represent the standard deviation.

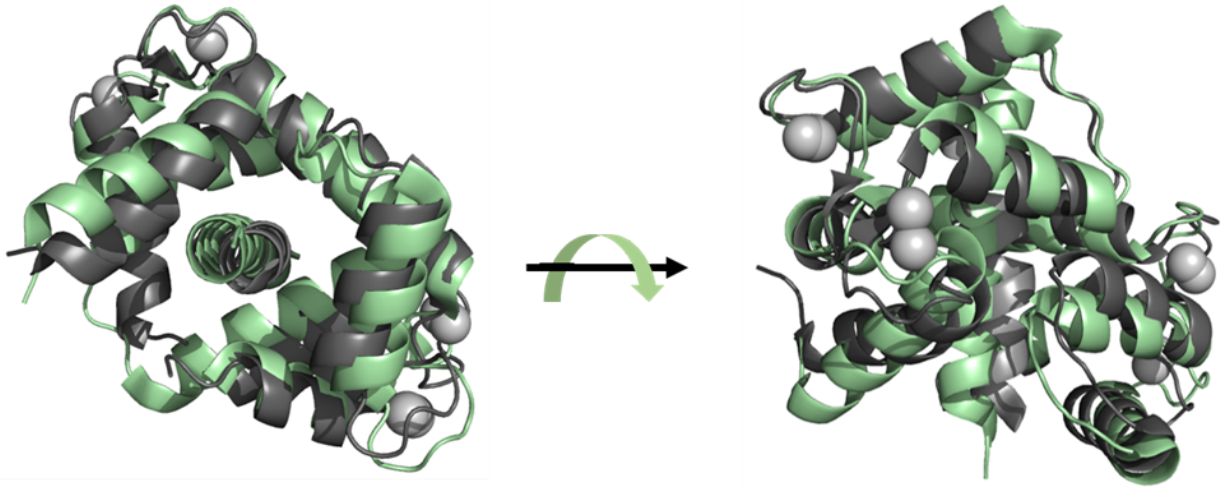

Figure 8. Comparison of CaM:pCaN (green) and CaM:pCaMKI (gray) complexes showing close similarities in the binding modes. RMSD = 2.6Å. CaM:pCaN from 4Q5U[15], and CaM:pCaMKI from 2L7L[23]

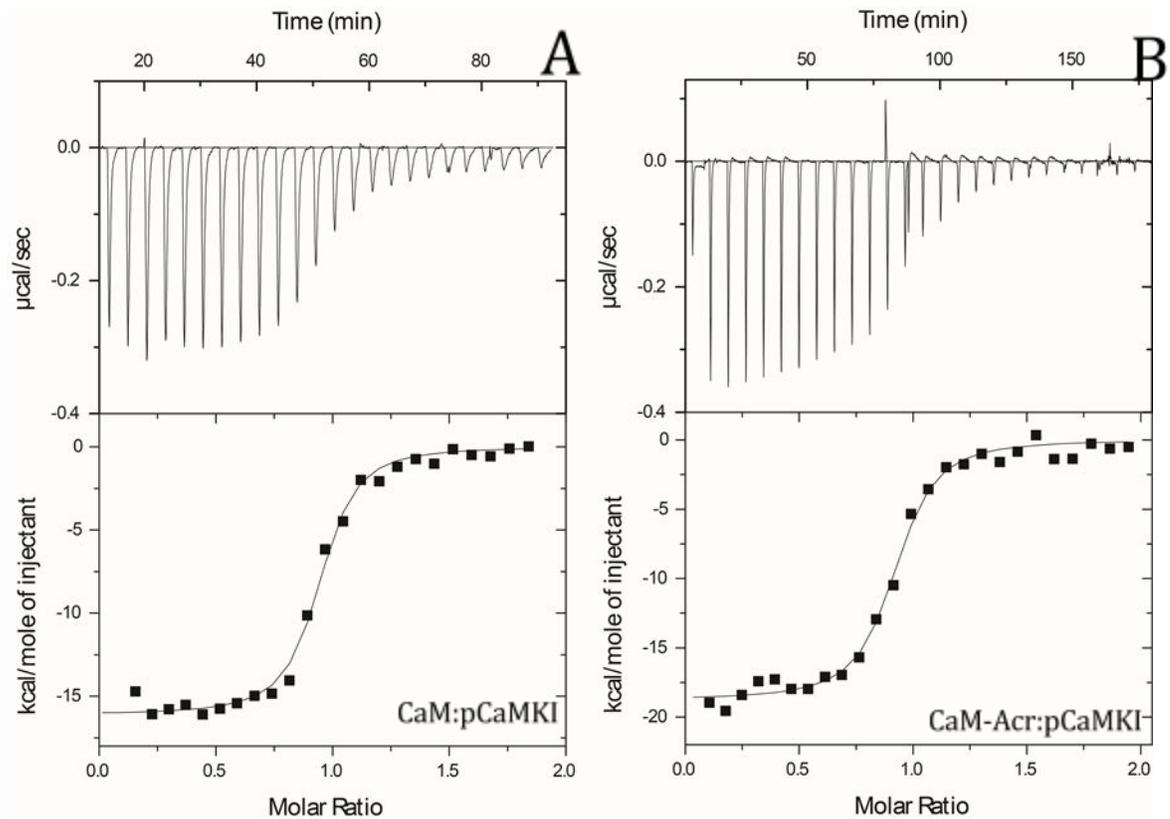

|     | CaM | CaM-Acr |
|-----|-----|---------|
| N:  | 0.917 | 0.888 |
|     | (±0.0078) | (±0.014) |
| K:  | 32.7 | 33.9 |
|     | (±5.2) | (±5.1) |
| ΔH: | -16.1 | -18.1 |
|     | (±0.2) | (±0.2) |
| ΔS: | -17.7 | -23.9 |

Figure 9. Representative ITC titrations of pCaMKI injected into either CaM (A) or CaM-Acr (B). A. 50μM pCaMKI injected into 5μM unlabeled CaM. B. 50μM pCaMKI injected into 5μM CaM D3C labeled with acrylodan (CaM-Acr). All ITC was performed under identical conditions and buffers. The jacket temperature was set to 37°C. Proteins and peptides were buffer exchanged into 20mM HEPES (pH 7.5), 250mM NaCl, and 10mM $CaCl_2$

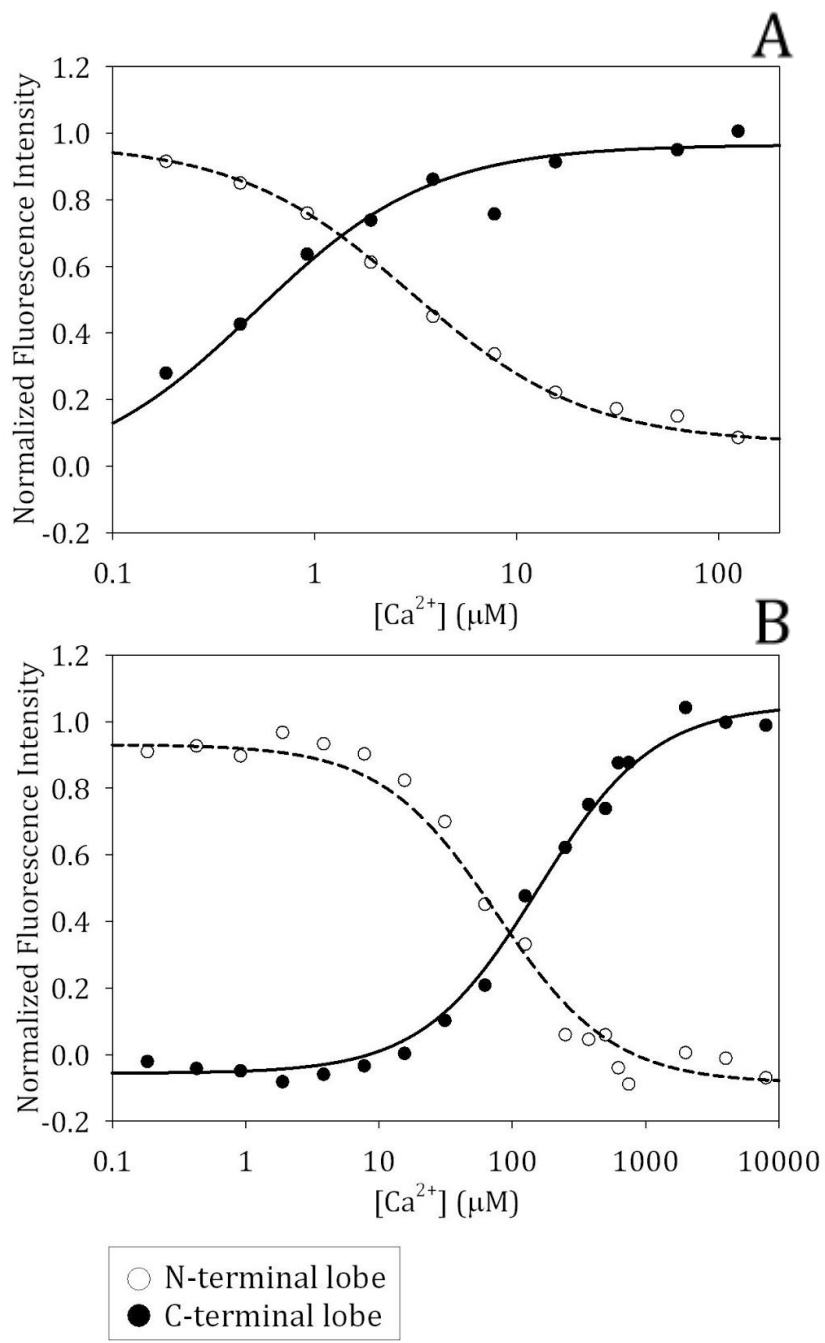

Figure 10. $Ca^{2+}$-binding affinities for the N-lobe and C-lobe of CaM in A) 20mM HEPES (pH 7.5) and 250mM NaCl and B) 20mM HEPES (pH 7.5) and 2M NaCl. The $Ca^{2+}$-binding affinity for both lobes decreases significantly due to electrostatic screening.

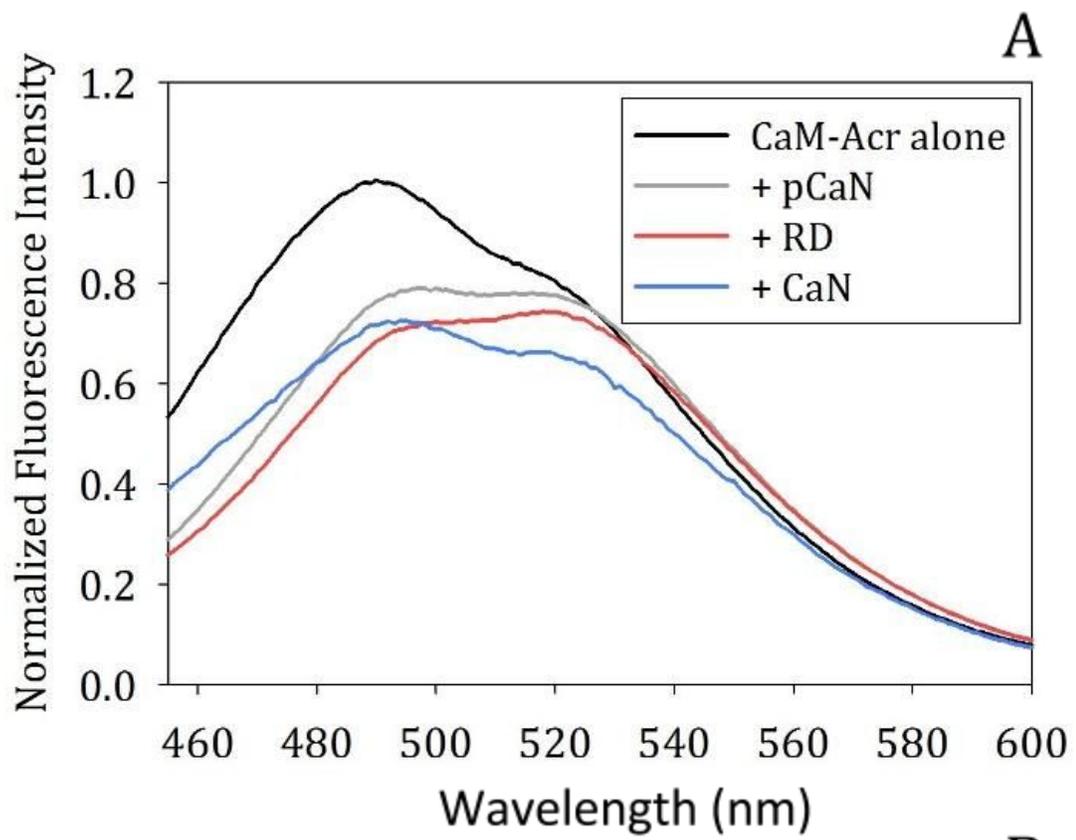

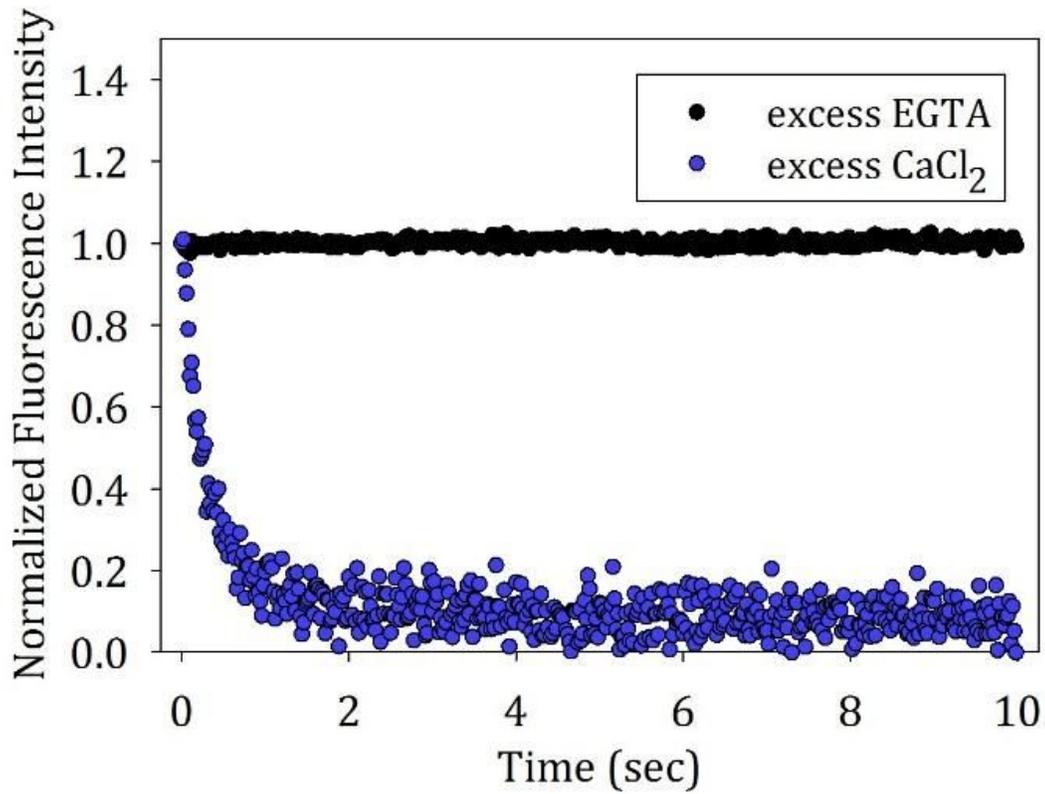

Figure 11. A. Fluorescent emission spectra of 0.2μM CaM-Acr in the presence and absence of 2μM pCaN, RD, and CaN at 37°C. Samples were excited at 365nm and contained 20mM HEPES, 250mM NaCl, and 10mM CaCl$_2$. B. The injection of 200μM CaN into 48μM CaM-Acr resulted in quenching of CaM-Acr fluorescence when 10mM CaCl$_2$ was present but not in the presence of 10mM EGTA.

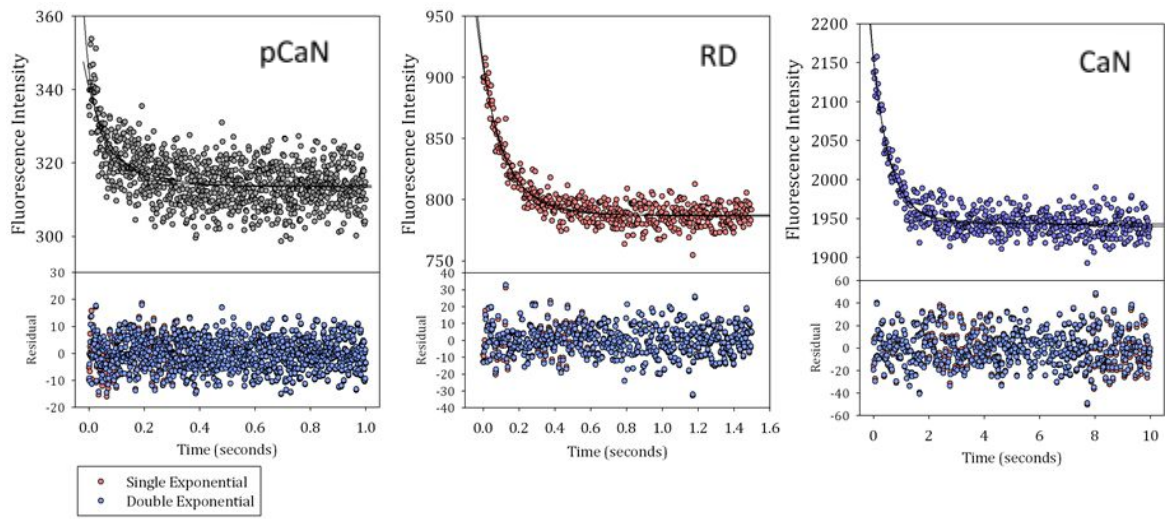

Figure 12. Representative association data for CaM-Acr binding to pCaN, RD, and CaN. Experimental data were fit to either a single (Equation 1) or double exponential decay (Equation 2). Both models resulted in randomly distributed residuals.

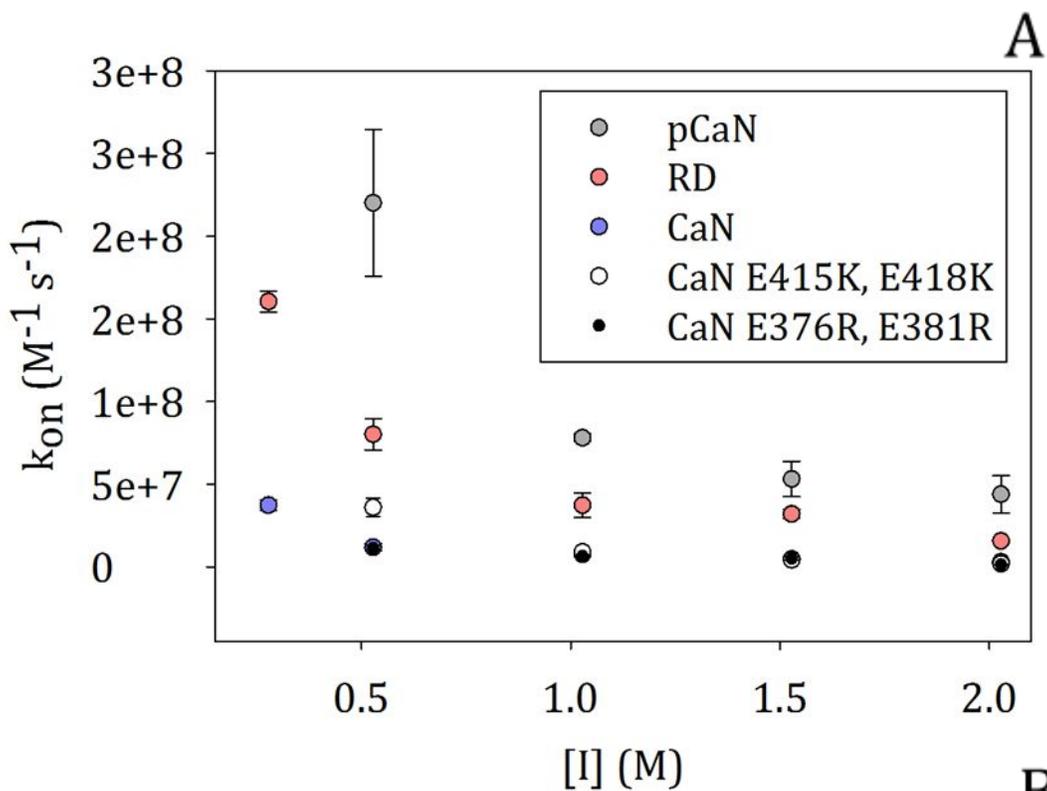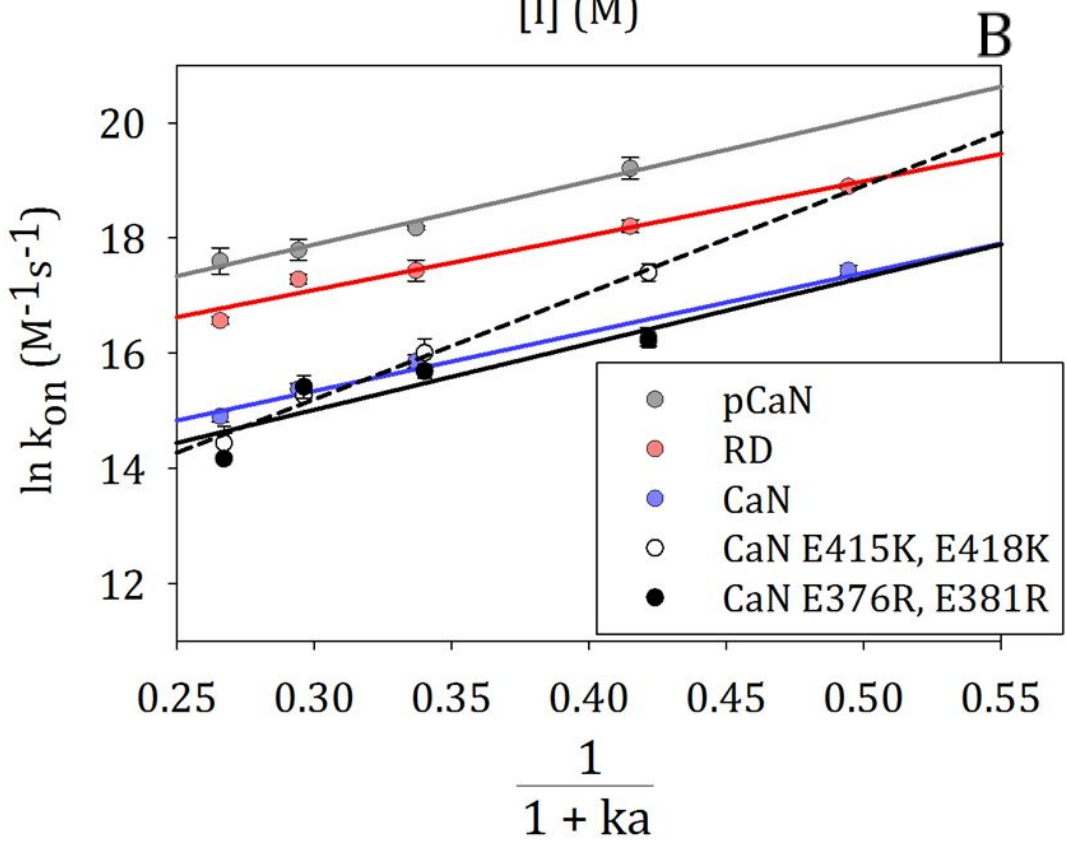

Figure 13. A) Association rate constants of CaM-Acr interacting with pCaN, RD, or CaN at varying ionic strengths B) Debye-Hückel plot of the association rate constants. Electrostatic energy of interaction (U) is calculated from the slope of ln $k_{on}$ vs. $1/1+\kappa a$

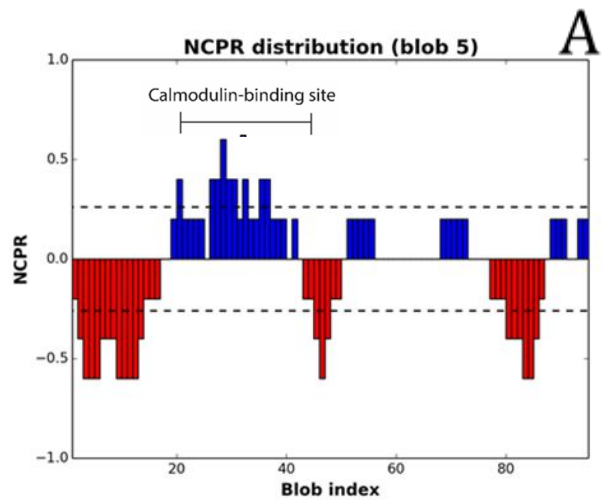
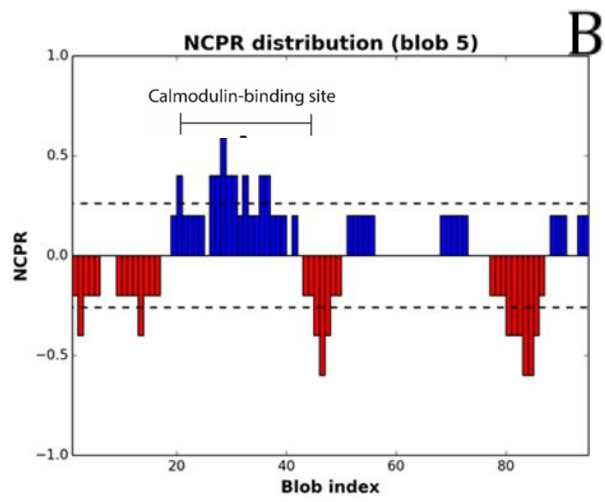
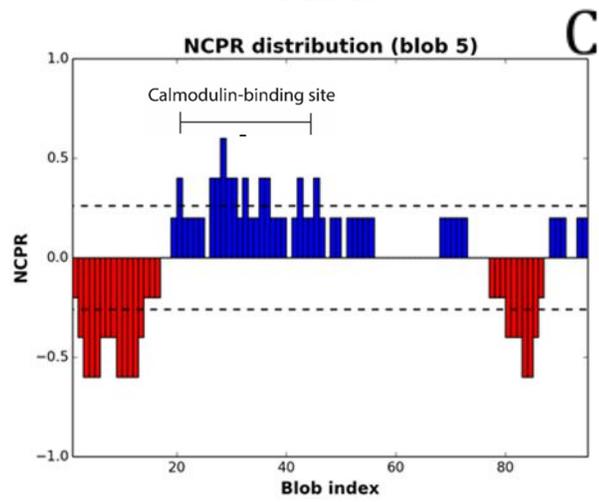

Figure 14. Net charge per residue (NCPR) of the RD (A), RD E376R, E381R (B), and RD E415K, E418K (C).

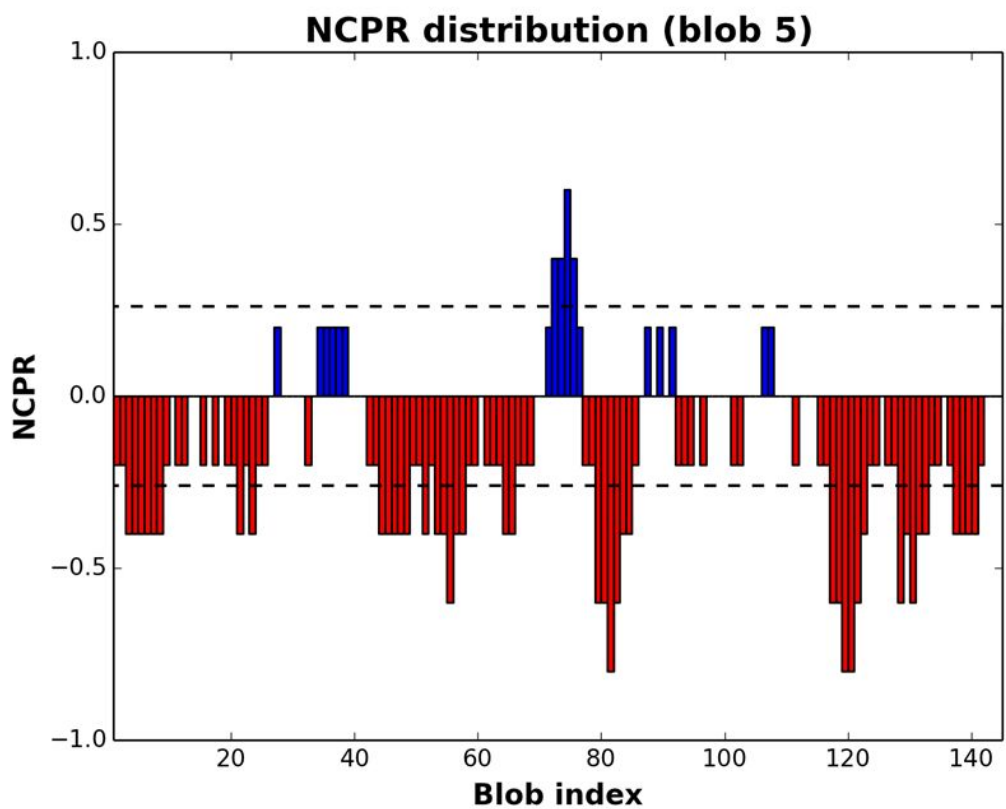

Figure 16. Linear net charge per residue plot of CaM using CIDER. The 75[th] blob index (also shown as a highly positively charged region in blue) represents the center linker connecting the two lobes of CaM.

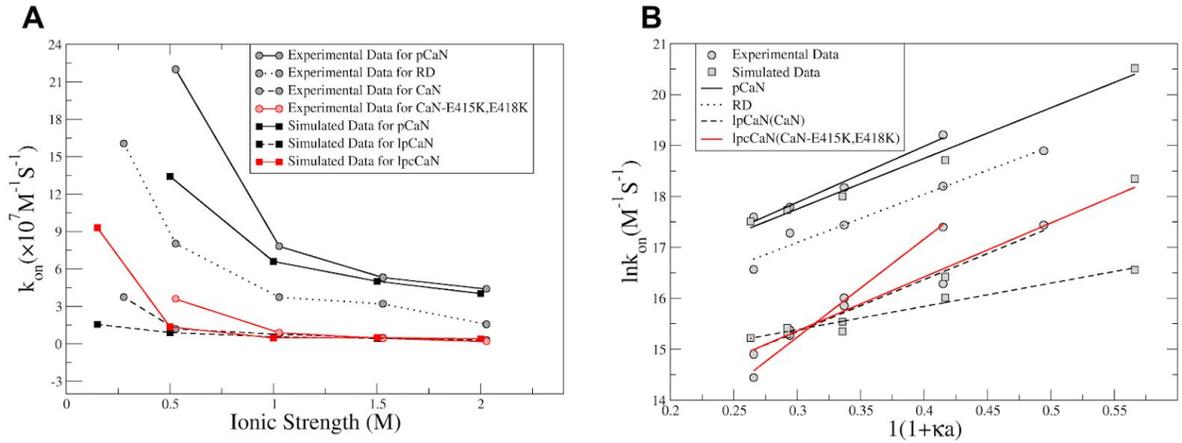

Figure 17. A) Comparison of simulated $k_{on}$ values for the CaM/CaN peptides with experimental measurements at varying ionic strengths. B) Comparison of Debye-Hückel plot of the association rate constants between computed results and experimental data.

| Protein/Peptide: | Net Charge at pH 7.5: |
|---|---|
| CaM - calmodulin | -16.2 |
| pCaN (peptide of CaM-binding region) | +6.7 |
| RD – regulatory domain | -3.7 |
| CaN – full-length calcineurin | -13.7 |
| CaN E376R, E381R | -9.7 |
| CaN E415K, E418K | -9.7 |

Table 1. Net charges of protein constructs at pH 7.5 were calculated using Protein Calculator v3.4 (protcalc.sourceforge.net). The net charges of CaM, CaN, and CaN E415K, E418K were increased by +8 to account for the $Ca^{2+}$-bound state of these molecules.

|  | $K_D$ at 0.25M NaCl (μM) | $K_D$ at 2.0M NaCl (μM) |
|---|---|---|
| CaM N-terminal lobe | 4.0 ± 0.5 | 77 ± 18 |
| CaM C-terminal lobe | 0.40 ± 0.05 | 160 ± 20 |

Table 2. The $Ca^{2+}$-binding affinity of CaM at 0.25M and 2.0M NaCl at 37°C. 20mM HEPES (pH 7.5) was present is all samples.

| CaM-substrate | Single Exponential $R^2$ | Double Exponential $R^2$ |
| --- | --- | --- |
| pCaN | 0.516 | 0.518 |
| RD | 0.970 | 0.970 |
| CaN | 0.959 | 0.964 |

Table 3. $R^2$ values for the fit of single and double exponential decay equations to the kinetics traces shown in Fig. 4.8. Single exponential decay model was chosen because double exponential models did not give a significantly better fit.

| CaM substrate | $k_{on,basal}$ (x $10^7$ $M^{-1}$ $s^{-1}$) | Energy of Electrostatic Interaction (U) (kcal $mol^{-1}$) |
|---|---|---|
| pCaN | 2.16 ± 0.05 | -6.7 ± 0.6 |
| RD | 1.5 ± 0.04 | -5.8 ± 0.6 |
| CaN | 0.21 ± 0.01 | -6.3 ± 0.8 |
| CaN E415K, E418K | 0.013 ± 0.0006 | -12 ± 0.8 |
| CaN E376R, E381R | 0.11 ± 0.01 | -7.1 ± 0.8 |

Table 4. Estimated energies of electrostatic interaction and $k_{on}$ basal values for the association of CaM-Acr with pCaN, RD, CaN, CaN E415K, E418K, and CaN E376R, E381R.

|  | (x $10^7$ M$^{-1}$ s$^{-1}$) | (U) (kcal mol$^{-1}$) |
|---|---|---|
| pCaN | 0.246 | -5.894 |
| lpCaN | 0.122 | -2.704 |
| lpcCaN | 0.019 | -6.278 |

Table 5. Computed energies of electrostatic interaction and $k_{on}$ basal values for the association of CaM with pCaN, lpCaN and lpcCaN peptides via Browndye.